\documentstyle[twoside,fleqn,espcrc2,epsf]{article}
\newcommand{\AmS}{{\protect\the\textfont2
  A\kern-.1667em\lower.5ex\hbox{M}\kern-.125emS}}

\title{
Large-$q$ expansion of the two-dimensional $q$-state Potts model
by the finite lattice method
}
\author{
H. Arisue
\address{Osaka Prefectural College of Technology, 
26-12 Saiwai-cho, Neyagawa, Osaka 572, Japan}
\thanks{This work is supported in part 
by the Grants-in-Aid of the Ministry of Education (No.08640494)
}
and  
K. Tabata
\address{Osaka Institute of Technology,
        Junior College, Ohmiya, Asahi-ku, Osaka 535, Japan}
}
       
\begin{document}

\begin{abstract}
We have calculated the large-$q$ expansion for the energy and magnetization 
cumulants at the first order phase transition point 
in the two-dimensional $q$-state Potts model to the 21st or 23rd 
order in $1/\sqrt{q}$ using the finite lattice method. The obtained series 
allow us to give highly convergent estimates of the cumulants for $q>4$. 
The results confirm us the correctness of 
the conjecture by Bhattacharya {\em et al.} on the asymptotic behavior of the 
energy cumulants for $q \rightarrow 4_+$ 
and a similar new conjecture on the magnetization cumulants. 
\end{abstract}

\maketitle

\section{INTRODUCTION}
The $q$-state Potts model\cite{Potts,Wu} in two dimensions has been 
investigated intensively 
as the test ground for analyzing the phase transition in
many physical systems. Especially it is interesting because the order of the 
phase transition changes when the parameter $q$ is varied, i.e.,
from the first order for $q>4$ 
to the second order for $q\le 4$.
The amplitudes of many quantities 
at the first order transition point are known exactly, including the 
latent heat, the spontaneous magnetization\cite{Baxter1973} 
and the correlation length\cite{Klumper,Buffenoir,Borgs}.
Bhattacharya {\em et al.}\cite{Bhattacharya1994} made a stimulating conjecture
on the asymptotic behavior of the energy cumulants (including the specific 
heat) at the first order transition point. 
Bhattacharya {\em et al.}\cite{Bhattacharya1997} also
made the large-$q$ expansion of the energy cumulants 
to order 10 in $z\equiv 1/\sqrt{q}$ and analyzing the expansion series
with the conjecture they could give the estimates of the cumulants
at the transition point for $q \ge 7$ 
that are better than those given by other methods 
including the Monte Carlo simulations\cite{Janke} and
the low-(and high-)temperature expansions\cite{Bhanot1993,Briggs,Arisue1997}.
The series obtained by Bhattacharya {\em et al.}
are, however, not long enough to investigate the behavior
of the energy cumulants for $q$ closer to $4$.

Here 
using the finite lattice method\cite{Enting1977,Creutz,Arisue1984}
we have generated the large-$q$ series for the energy cumulants
and the magnetization cumulants
at the transition point to order 21 or 23 in $z$\cite{Arisue1998}.
The finite lattice method can in general give longer series 
than those obtained by the diagrammatic method 
especially in lower space (and time) dimensions. 
In the diagrammatic method, 
one has to list up all the relevant diagrams and count the number they appear.
In the finite lattice method we can skip this job and 
reduce the main work to the calculation of the partition function 
for a series of finite size lattices,
which can be done without the diagrammatic technique.
This method has been used mainly to 
generate the low- and high-temperature series in statistical systems and
the strong coupling series in lattice gauge theory.
One of the purposes of our work is to demonstrate that this method 
is also applicable to the series expansion with respect to the parameter 
other than the temperature or the coupling constant.

\section{ENERGY CUMULANTS}
The latent heat ${\cal L}$ 
and the correlation length $\xi$ 
at the transition point are known to vanish and diverge, respectively,
at $q\rightarrow 4_+$ as
\begin{eqnarray}
{\cal L} &\sim& 3\pi x^{-1/2}\;, \label{eq:latentheat} \\
\xi      &\sim& \frac{1}{8\sqrt{2}} x \label{eq:correlation}
\end{eqnarray}
with $x=\exp{(\pi^2/2\theta)}$ and $2\cosh{\theta}=\sqrt{q}$.
Bhattacharya {\em et al.}'s conjecture says
that the $n$-th energy cumulants $F_{d,o}^{(n)}$ 
at the first order transition point $\beta=\beta_t$ will 
diverge at $q\rightarrow 4_+$ as
\begin{equation}
F_d^{(n)}, (-1)^n F_o^{(n)} \sim \alpha B^{n-2}
  \frac{\Gamma\left(n-\frac{4}{3}\right)}{\Gamma
    \left(\frac{2}{3}\right)}x^{3n/2-2}\;.\label{eq:asymp_form}
\end{equation}
This is from the fact that 
for $\beta\rightarrow\beta_t$ at $q=4$ (the second order phase transition)
the correlation length and the second cumulants diverge as 
$\xi \sim \lambda |\beta-\beta_t|^{-2/3}$ and 
$F^{(2)} \sim \mu |\beta-\beta_t|^{-2/3}$, respectively, so that 
$F^{(n)} \sim \mu\frac{\Gamma\left(n-\frac{4}{3}\right)}
{\Gamma\left(\frac{2}{3}\right)}(\xi/\lambda)^{3n/2-2}$ 
and from the assumption that this relation between 
the correlation length and the energy cumulants are also kept
for $q\rightarrow 4_+$ with $\beta=\beta_t$.
The constants $\alpha$ and $B$ in Eq.(\ref{eq:asymp_form}) 
should be common to the ordered and disordered phases
from the duality relation for each $n$-th cumulants.

\begin{table}[tb]
\caption{
The specific heat for some values of $q$.
The exact correlation length is also listed.
         }
\begin{center}
\begin{tabular}{rlll}
\hline
    $q$  & \multicolumn{1}{c}{$C_d$}
         & \multicolumn{1}{c}{$C_o$}  
         & \multicolumn{1}{c}{$\xi_d$} \\
\hline
$  5$ & 2889(2)         & 2886(3)         & 2512.2 \\ 
$  6$ & 205.93(3)       & 205.78(3)       & 158.9 \\ 
$  7$ & 68.738(2)       & 68.513(2)       & 48.1 \\ 
$  8$ & 36.9335(3)      & 36.6235(3)      & 23.9 \\ 
$  9$ & 24.58761(8)     & 24.20344(7)     & 14.9 \\ 
$ 10$ & 18.38543(2)     & 17.93780(2)     & 10.6 \\ 
$ 12$ & 12.401336(3)    & 11.852175(2)    & 6.5 \\ 
$ 15$ & 8.6540358(4)    & 7.9964587(2)    & 4.2 \\ 
$ 20$ & 6.13215967(2)   & 5.36076877(1)   & 2.7 \\ 
\hline
\end{tabular}
\end{center}
\end{table}
\begin{figure}[tb]
\epsfxsize=7.4cm
\epsffile{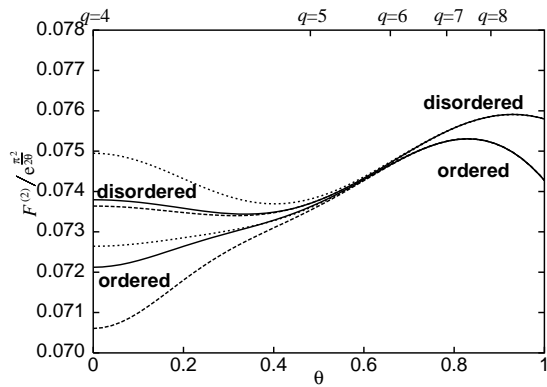}
\vspace{-0.7cm}
\caption{
The behavior of the ratio of the second energy cumulants 
$F_{d,o}^{(2)}$ to $x$
         }
\vspace{-0.4cm}
\end{figure}
If this conjecture is true, 
the product $F^{(n)}{\cal L}^{3n-4}$ is a smooth function of $\theta$, 
so we can expect that the Pad\'e approximation of $F^{(n)}{\cal L}^p$ 
will give convergent result at $p=3n-4$.
It has been tried for the large-$q$ series obtained by the finite lattice 
methods for $n=2,\cdots,6$ both in the ordered and disordered phases, 
which in fact gives quite convergent Pad\'e approximants 
for $p=3n-4$ and as $p$ leaves from this value the convergence of the 
approximants becomes bad rapidly. 
We give in Table 1 the values of the 
specific heat $C=\beta_t^2F^{(2)}$ evaluated 
from these Pad\'e approximants for some values of $q$ 
and present in Fig.1 the behavior of the ratio $F^{(2)}/x$ 
plotted versus $\theta$.
These estimates are three or four orders of magnitude more precise than
(and consistent with)
the previous result for $q\ge 7$ 
from the large-$q$ expansion to order $z^{10}$
by Bhattacharya {\em et al.}\cite{Bhattacharya1997} and 
the result of the Monte Carlo simulations for $q=10,15,20$
carefully done by Janke and Kappler\cite{Janke}. 
What should be emphasized is that we obtained the values of
the specific heat in the accuracy of about 0.1 percent
at $q=5$ where the correlation length is as large as 2500.
As for the asymptotic behavior of $F^{(n)}$ at $q\rightarrow 4_+$,
the Pad\'e data of $F_d^{(2)}/x$ and $F_o^{(2)}/x$  shown in Fig.2
have the errors of a few percent around $q=4$ and 
their behaviors are enough to convince us 
that the conjecture (\ref{eq:asymp_form}) is true for $n=2$ with
$\alpha = 0.073 \pm 0.002\;.$
Furthermore from the conjecture (\ref{eq:asymp_form}) the combination 
$\left\{{\Gamma\left(n-\frac{4}{3}\right)|F^{(n)}|}/
{\Gamma\left(\frac{2}{3}\right)F^{(2)}}\right\}^{\frac{1}{n-2}}
x^{-\frac{3}{2}}$ is expected 
to approach the constant $B$ for each $n(\ge 3)$, 
and in fact the Pad\'e data for every $n(=3,\cdots,6)$ gives 
$B=0.38\pm 0.05\;,$ 
which gives strong support to the conjecture also for $n\ge 3$.

\section{MAGNETIZATION CUMULANTS}
The behavior of the $n$-th magnetization cumulants $M^{(n)}$ 
for $\beta\to\beta_t$ at $q=4$ is well known as
$M_{d,o}^{(n)} \simeq A_{d,o}^{(n)} (\xi)^{\frac{15}{8}n-2}$
and we can make a conjecture parallel to the conjecture 
for the energy cumulants by Bhattacharya {\it et al.}: 
this relation will also be kept in the limit $q\to 4_+$ with $\beta=\beta_t$, 
which implies 
using the asymptotic behavior of the correlation length 
in Eq.(\ref{eq:correlation})
in this limit that
\begin{equation}
M_{d,o}^{(n)} \sim C_{d,o}^{(n)} x^{\frac{15}{8}n-2}\;.\label{eq:asymp_m}
\end{equation}
We have tried the Pad\'e approximation of $M^{(n)}{\cal{L}}^{p}$ 
for the large-$q$ series generated 
by the finite lattice method to order $z^{21}$
for $n=2$ and $3$ both in the ordered and disordered phases, 
which in fact gives quite convergent Pad\'e approximants 
for $p=15n/4-4$
and as $p$ leaves from this value the convergence of the 
approximants becomes bad rapidly again. 
In Table 2 we present the resulting estimates of the magnetic susceptibility
$\chi_{d,o}=M_{d,o}^{(2)}$.
Our result is much more precise than the Monte Carlo simulation\cite{Janke} 
at least by a factor of 100.
From the behavior of $M_{d,o}^{(n)}/x^{(15n/8-2)}$ 
we obtain the coefficients in Eq.(\ref{eq:asymp_m}) as 
$C_{d}^{(2)}=0.0020(2)$, $C_{o}^{(2)}=0.0016(1)$ 
and $C_{d}^{(3)}=7.4(5)\times 10^{-5}$, $C_{o}^{(3)}=7.9(2)\times 10^{-5}$. 
These convince us that the conjecture made for the magnetization cumulants 
is also true.

\begin{table}[t]
\caption{
The magnetic susceptibility for some values of $q$.
         }
\begin{center}
\begin{tabular}{rll}
\hline
$q$  & \multicolumn{1}{c}{$\chi_d$} 
     & \multicolumn{1}{c}{$\chi_o$} \\
\hline
$  5$ & $9.13(3)\times 10^4 $ & $9.01(3)\times 10^4  $ \\ 
$  6$ & $6.585(4)\times 10^2$ & $6.665(4)\times 10^2 $ \\ 
$  7$ & $70.54(1)           $ & $77.31(1)            $ \\ 
$  8$ & $19.359(1)          $ & $21.525(1)           $ \\ 
$  9$ & $8.0579(1)          $ & $9.0106(2)          $ \\ 
$ 10$ & $4.23276(2)         $ & $4.73823(4)         $ \\ 
$ 12$ & $1.720645(2)        $ & $1.918274(3)        $ \\ 
$ 15$ & $0.7304214(1)       $ & $0.8056969(2)       $ \\ 
$ 20$ & $0.309365682(1)     $ & $0.33556421(1)      $ \\ 
\hline\end{tabular}
\end{center}
\end{table}

\section{SUMMARY}
The large-$q$ series for the energy and magnetization cumulants 
generated by the finite lattice method 
in the two-dimensional $q$-state Potts model
give very precise estimates of the cumulants for $q>4$
and they confirm the correctness of the conjecture that 
the relation between the cumulants and the correlation length 
for $q=4$ and $\beta\to\beta_t$ (the second order phase transition)
is kept in their asymptotic behavior for $q\to 4_+$ 
at $\beta=\beta_t$ (the first order transition point).
If this kind of relation is satisfied 
for the quantities at the first order phase transition point
in more general systems as the asymptotic behavior 
when the parameter of the system is varied 
to make the system close to the second order phase transition point,
it would serve as a good guide in investigating the property of 
the system.



\begin{thebibliography}{9}
\bibitem{Potts}
  R. B. Potts,
     Proc. Camb. Phil. Soc. {\bf 48}, 106 (1952).
\bibitem{Wu}
  F. Y. Wu, 
     Rev. Mod. Phys. {\bf 54}, 235 (1982).
\bibitem{Baxter1973}
  R. J. Baxter, 
     J. Phys. C {\bf 6},  L445 (1973);
     J. Stat. Phys. {\bf 9}, 145 (1973). 
\bibitem{Klumper}
  A. Kl\"{u}mper, A. Schadschneider and J. Zittartz, 
     Z. Phys. B {\bf 76}, 247 (1989).
\bibitem{Buffenoir}
  E. Buffenoir and S. Wallon,
     J. Phys. A {\bf 26}, 3045 (1993).
\bibitem{Borgs}
  C. Borgs and W. Janke, 
     J. Phys. I (France) {\bf 2}, 649 (1992).
\bibitem{Bhattacharya1994}
   T.~ Bhattacharya, R.~Lacaze and A.~Morel,
     Nucl. Phys. {\bf B435}, 526 (1995).
\bibitem{Bhattacharya1997}
   T.~ Bhattacharya, R.~Lacaze and A.~Morel,
      J. Phys. I (France) {\bf 7}, 1155 (1997).
\bibitem{Janke}
  W. Janke and S. Kappler,
     J. Phys. I (France) {\bf 7}, 1155 (1997).
\bibitem{Bhanot1993}
  G. Bhanot, M. Creutz, U. Gl\"assner, I. Horvath , J. Lacki,
  K. Schilling  and J. Weckel, 
     Phys. Rev. B {\bf 48}, 6183 (1993).
\bibitem{Briggs}
  K. M. Briggs, I. G. Enting and A. J. Guttmann, 
     J. Phys. A {\bf 27}, 1503 (1994).       
\bibitem{Arisue1997}
  H. Arisue and K. Tabata,
  J. Phys. A {\bf 30}, 3313 (1997). 
\bibitem{Enting1977}
  T. de Neef and I. G. Enting, 
     J. Phys. A {\bf 10}, 801 (1977);
  I. G. Enting,
     J. Phys. A {\bf 11}, 563 (1978);
     Nucl. Phys. B (Proc. Suppl.) {\bf 47}, 180 (1996).
\bibitem{Creutz}
  M. Creutz, 
     Phys. Rev. B {\bf 43}, 10659 (1991).
\bibitem{Arisue1984}
    H. Arisue and T. Fujiwara, 
      Prog. Theor. Phys. {\bf 72}, 1176 (1984);
    H. Arisue,
      Nucl. Phys. B (Proc. Suppl.) {\bf 34}, 240 (1994).
\bibitem{Arisue1998}
  H. Arisue and K. Tabata,
  preprint hep-lat/9807005 and in preparation.
\end{thebibliography}
\end{document}